\documentclass{article}
\usepackage{amssymb}
\usepackage{amsfonts}
\usepackage{amsmath}

\newtheorem{theorem}{Theorem}
\newtheorem{acknowledgement}[theorem]{Acknowledgement}

\newtheorem{conclusion}[theorem]{Conclusion}

\input{tcilatex}
\begin{document}

\title{Can Anyon statistics explain high temperature superconductivity?}
\author{Ahmad Adel Abutaleb \\
Department of Mathematics, Faculty of science,\\
University of \ \ Mansoura, Elmansoura 35516, Egypt}
\maketitle

\begin{abstract}
In this paper, we find a reasonable explanation of high temperature
superconductivity phenomena using Anyon statistics.
\end{abstract}

\section{\textbf{Introduction}}

Superconductivity is a phenomena occurring in certain materials when cooled
below some characteristic temperature called a critical temperature. In a
superconductor, the resistance drops suddenly to zero when the material is
cooled below its critical temperature. This phenomena was first discovered
by the dutch physicist Heike Kamerlingh Onnes, which received the Nobel
prize of physics in 1913 for discovering this phenomena and the related work
regarding liquefaction of helium [1]. There is another phenomena related to
superconductor called (Meissner effect), which is the expulsion of a
magnetic field from a superconductor during its transition to the
superconducting state. This phenomena was discovered in 1933 by Walther
Meissner and Robert Ochsenfeld [2]. Although the microscopic theory of
superconductivity was developed by Bardeen, Cooper and Schrieffer (BCS
theory) in 1957 [3], other descriptions of the phenomena of the
superconductivity were developed [4-6].

Whereas ordinary superconductors usually have transition temperatures below
30 Kelvin, It is well known that some materials behave as superconductors at
unusually high critical temperatures and this materials called high
temperature superconductors or (HTS). The first HTS was discovered in 1986
by Georg Bednorz and K. Alex Muller, who were awarded the 1987 Nobel Prize
in Physics [7].

Although the standard BCS theory (weak-coupling electron--phonon
interaction) can explain all known low temperature superconductors and even
some high temperature superconductors, some of high temperature
superconductors cannot be explained by BCS theory and some physicist believe
that they obey a strong interaction.

In this paper, we find that if we use a unified statistics [8] instead
Fermi-Dirac statistics, then we can find a reasonable explanation of all
high temperature superconductivity phenomena using a standard BCS theory.

\section{\textbf{The model}}

We follow the derivation of Hamiltonian model using the mean field
approximation and Bogoliubov unitary transformation as in the reference [9],
where we can write the following relation,

\begin{equation}
\left\vert \Delta \right\vert =\left\vert V\right\vert \sum_{k}\frac{%
\left\vert \Delta \right\vert }{2\sqrt{(\epsilon (k)-\mu )^{2}+\Delta ^{2}}}%
\left[ f(-E_{K})-f(E_{K})\right] \text{,}
\end{equation}

where $f(E_{K})$ determined from the unified statistics as follow [8],

\begin{equation}
\left( \frac{1}{(1-p)f(E_{k})}+1\right) ^{1-p}\left( \frac{1}{pf(E_{k})}%
-1\right) ^{p}=e^{\dfrac{\sqrt{(\epsilon (k)-\mu )^{2}+\Delta ^{2}}}{k_{B}T}}%
\text{.}
\end{equation}

\section{\textbf{BCS Theory ( }$p\rightarrow 1)$}

\begin{center}
\textbf{For }$p\rightarrow 1$\textbf{, we recover the usual BCS Theory.}
\end{center}

When $p\rightarrow 1,$we have the following usual form of the Fermi function 
$f(E_{K})$,

\begin{equation}
f(E_{K})=\frac{1}{1+e^{\dfrac{\sqrt{(\epsilon -\mu )^{2}+\Delta ^{2}}}{k_{B}T%
}}}\text{,}
\end{equation}

so, we have,

\begin{equation}
1=\left\vert V\right\vert \sum_{k}\int \frac{1}{2\sqrt{(\epsilon (k)-\mu
)^{2}+\Delta ^{2}}}\left[ \frac{1}{1+e^{\dfrac{-\sqrt{(\epsilon (k)-\mu
)^{2}+\Delta ^{2}}}{k_{B}T}}}-\frac{1}{1+e^{\dfrac{\sqrt{(\epsilon (k)-\mu
)^{2}+\Delta ^{2}}}{k_{B}T}}}\right] \text{.}
\end{equation}

Replacing the summation by integration over the energy states, we have,

\begin{equation}
1=\left\vert V\right\vert \int_{\mu -\epsilon _{D}}^{\mu +\epsilon _{D}}%
\frac{N(0)d\epsilon }{2\sqrt{(\epsilon -\mu )^{2}+\Delta ^{2}}}\left[ \frac{1%
}{1+e^{\dfrac{-\sqrt{(\epsilon -\mu )^{2}+\Delta ^{2}}}{k_{B}T}}}-\frac{1}{%
1+e^{\dfrac{\sqrt{(\epsilon -\mu )^{2}+\Delta ^{2}}}{k_{B}T}}}\right] \text{.%
}
\end{equation}

Let $\zeta =\epsilon -\mu ,$ then we have the master equation for BCS theory
as,

\begin{equation}
\frac{1}{\left\vert V\right\vert N(0)}=\int_{0}^{\epsilon _{D}}\dfrac{d\zeta 
}{\sqrt{\zeta ^{2}+\Delta ^{2}}}\left[ \frac{1}{1+e^{\dfrac{-\sqrt{\zeta
^{2}+\Delta ^{2}}}{k_{B}T}}}-\frac{1}{1+e^{\dfrac{\sqrt{\zeta ^{2}+\Delta
^{2}}}{k_{B}T}}}\right] \text{.}
\end{equation}

\begin{center}
\textbf{Determine }$T_{C}$\textbf{.}
\end{center}

At $\Delta =0$, we have $T=T_{C}$, so we can write,

\begin{equation}
\frac{1}{\left\vert V\right\vert N(0)}=\int_{0}^{\epsilon _{D}}\dfrac{d\zeta 
}{\zeta }\left[ \frac{1}{1+e^{\dfrac{-\zeta }{k_{B}T_{C}}}}-\frac{1}{1+e^{%
\dfrac{\zeta }{k_{B}T_{C}}}}\right] \text{.}
\end{equation}

Define $\dfrac{\zeta }{k_{B}T_{C}}=x$, we can write,

\begin{equation}
\frac{1}{\left\vert V\right\vert N(0)}=\int_{0}^{\dfrac{{\huge \epsilon }_{D}%
}{K_{B}T_{C}}}\dfrac{dx}{x}\phi (x)\text{,}
\end{equation}

where $\phi (x)=\left[ \dfrac{1}{1+e^{-x}}-\dfrac{1}{1+e^{x}}\right] $.

Using integration by parts we have,

\begin{equation}
\int_{0}^{\dfrac{{\Huge \epsilon }_{D}}{K_{B}T_{C}}}\dfrac{dx}{x}\phi (x)=%
\left[ \phi (x)\ln x\right] _{0}^{\dfrac{{\huge \epsilon }_{D}}{K_{B}T_{C}}%
}-\int_{0}^{\dfrac{{\huge \epsilon }_{D}}{K_{B}T_{C}}}\frac{d\phi (x)}{dx}%
\ln xdx\text{.}
\end{equation}

For a weak coupling limit, i.e. $\left\{ \frac{1}{\left\vert V\right\vert
N_{0}}\gtrdot \gtrdot 1,k_{B}T_{C}\lessdot \lessdot \epsilon _{D},\left\vert
\Delta _{0}\right\vert \lessdot \lessdot k_{B}T_{C}\right\} ,$we have,

\begin{eqnarray}
\left[ \phi (x)\ln x\right] _{0}^{\dfrac{{\huge \epsilon }_{D}}{K_{B}T_{C}}}
&=&\ln \frac{\epsilon _{D}}{K_{B}T_{C}}\text{,} \\
\int_{0}^{\dfrac{{\huge \epsilon }_{D}}{K_{B}T_{C}}}\dfrac{d\phi (x)}{dx}\ln
xdx &\approx &\int_{0}^{\infty }\frac{d\phi (x)}{dx}\ln xdx\approx -0,12563%
\text{.}
\end{eqnarray}

So, we have

\begin{equation}
\frac{1}{\left\vert V\right\vert N(0)}=\ln \left( 1,134\frac{\epsilon _{D}}{%
K_{B}T_{C}}\right) \text{.}
\end{equation}

We can rewrite the last equation to obtain the well known formula for
critical temperature for BCS Theory as,

\begin{equation}
T_{C}=\frac{1,134\epsilon _{D}}{K_{B}}e^{\dfrac{-1}{\left\vert V\right\vert
N(0)}}\text{.}
\end{equation}

\begin{center}
\textbf{Determine }$\Delta _{0}.$
\end{center}

At $T=0$, we have $\Delta =\Delta _{0}$, so from the equation we have,

\begin{equation}
\frac{1}{\left\vert V\right\vert N(0)}=\int_{0}^{\epsilon _{D}}\dfrac{d\zeta 
}{\sqrt{\zeta ^{2}+\Delta _{0}^{2}}}=\sinh ^{-1}\left( \frac{\epsilon _{D}}{%
\Delta _{0}}\right) \text{,}
\end{equation}

so we have,

\begin{equation}
\frac{\epsilon _{D}}{\Delta _{0}}=\sinh \left( \frac{1}{\left\vert
V\right\vert N(0)}\right) \text{.}
\end{equation}

Again, for a weak coupling limit, we have $\sinh \left( \frac{1}{\left\vert
V\right\vert N(0)}\right) \approx \frac{1}{2}e^{\dfrac{1}{\left\vert
V\right\vert N(0)}}$, so we can write the BCS formula of $\Delta _{0}$ as
follow,

\begin{equation}
\frac{\epsilon _{D}}{\Delta _{0}}=\frac{1}{2}e^{\frac{1}{\left\vert
V\right\vert N(0)}}\text{.}
\end{equation}

From equations (13) and (16), we arrive at the universal BCS formula,

\begin{equation}
\Delta _{0}=1.76K_{B}T_{C}\text{.}
\end{equation}

Using equations (16) and (17) in the master equation (6), we can plot the
universal curve $y=y(x)$, where $x=\dfrac{T}{T_{C}},y=\dfrac{\Delta }{\Delta
_{0}}$ as follow,

\begin{equation}
\frac{1}{\left\vert V\right\vert N(0)}=\int_{0}^{\dfrac{\sinh \frac{1}{%
\left\vert V\right\vert N(0)}}{y}}\frac{dz}{\sqrt{z^{2}+1}}\left( \frac{1}{%
1+e^{\dfrac{-1.76y\sqrt{z^{2}+1}}{x}}}-\frac{1}{1+e^{\dfrac{1.76y\sqrt{%
z^{2}+1}}{x}}}\right) \text{.}
\end{equation}

\section{Anyon superconductivity}

From equations (1) and (2), we can write the general form of the master
equation for any value of $p$ as follow,

\begin{equation}
\frac{1}{\left\vert V\right\vert N(0)}=\dint\limits_{0}^{{\Huge \epsilon }%
_{D}}\frac{d\zeta }{\sqrt{\zeta ^{2}+\Delta ^{2}}}\left[ \phi (\frac{1}{%
r(\zeta )})-\phi (r(\zeta ))\right] \text{,}
\end{equation}

where,%
\begin{equation}
r(\zeta )=e^{^{\dfrac{\sqrt{\zeta ^{2}+\Delta ^{2}}}{K_{B}T}}\text{,}}
\end{equation}

and $\phi (r)$ determined from the following equation,

\QTP{Body Math}
\begin{equation}
\left( \frac{1}{(1-p)\phi (r)}+1\right) ^{1-p}\left( \frac{1}{p\phi (r)}%
-1\right) ^{p}=r\text{.}
\end{equation}

\subsection{\textbf{Determine }$T_{C}$\textbf{\ for any value of }$p$}

At $\Delta =0$, we have $T=T_{C}$ and then from equations (19-21) we can
write,

\QTP{Body Math}
\begin{eqnarray}
\frac{1}{\left\vert V\right\vert N(0)} &=&\dint\limits_{0}^{\dfrac{{\Huge %
\epsilon }_{D}}{K_{B}T_{C}}}\frac{dx}{x}\left[ \phi (\frac{1}{r(x)})-\phi
(r(x))\right] \text{,} \\
r(x) &=&e^{x},
\end{eqnarray}

and again, $\phi (r)$ determined from the equation (21).

\subsection{\textbf{Determine }$\Delta _{0}$ \textbf{for any value of }$p$}

At $T=0$, we have $\Delta =\Delta _{0}$ and from equation (20), we have $%
r=\infty $, so equation (21) takes the form,

\begin{equation}
\left( \frac{1}{(1-p)\phi (r)}+1\right) ^{1-p}\left( \frac{1}{p\phi (r)}%
-1\right) ^{p}=\infty \text{.}
\end{equation}

Then for any value of $p,$we have

\begin{equation}
\phi (r)=0\text{ at }T=0.
\end{equation}

Similarly, we can evaluate $\phi \left( \frac{1}{r}\right) $ by solving the
following equation,

\begin{equation}
\left( \frac{1}{(1-p)\phi \left( \dfrac{1}{r}\right) }+1\right) ^{1-p}\left( 
\frac{1}{p\phi \left( \dfrac{1}{r}\right) }-1\right) ^{p}=\frac{1}{r}\text{.}
\end{equation}

So at $T=0,$ We have $r=\infty ,\dfrac{1}{r}=0$ and then equation (26) takes
the form,

\begin{equation}
\left( \frac{1}{(1-p)\phi \left( \dfrac{1}{r}\right) }+1\right) ^{1-p}\left( 
\frac{1}{p\phi \left( \dfrac{1}{r}\right) }-1\right) ^{p}=0
\end{equation}

The solution of (26) (for positive $\phi (\frac{1}{r})$) takes the form,

\begin{equation}
\phi (\frac{1}{r})=\frac{1}{p}\text{ at }T=0\text{.}
\end{equation}

Using (25) and (28) in the master equation (19), we have at $T=0$ the
following equation,

\begin{equation}
\frac{p}{\left\vert V\right\vert N(0)}=\dint\limits_{0}^{{\Huge \epsilon }%
_{D}}\frac{d\zeta }{\sqrt{\zeta ^{2}+\Delta _{0}^{2}}}=\sinh ^{-1}\dfrac{%
\epsilon _{D}}{\Delta _{0}}\text{,}
\end{equation}

or in the equivalent form,

\begin{equation}
\epsilon _{D}=\Delta _{0}\sinh \frac{p}{\left\vert V\right\vert N(0)}\text{.}
\end{equation}

At weak coupling limit, from the equation (30), we have the following
universal relation for $\Delta _{0}$ at any value of $p$ as,

\begin{equation}
\Delta _{0}=2\epsilon _{D}\text{ }e^{\dfrac{-p}{\left\vert V\right\vert N(0)}%
}\text{.}
\end{equation}

In the next section, we will study some special cases of Anyon
superconductivity for some special values of $p$.

\bigskip

\section{Special cases of Anyon superconductivity}

\subsection{$p=\frac{1}{2}$ Anyon}

From the unified statistics (equation(2)), we can find the formula of $%
f(E_{k})$ as follow,

\begin{equation}
f(E_{k})=\frac{2}{\left( 1+e^{\dfrac{2\sqrt{(\epsilon -\mu )^{2}+\Delta ^{2}}%
}{k_{B}T}}\right) ^{\dfrac{1}{2}}}\text{.}
\end{equation}

So, after some calculations, we arrive at the master equation for the case $%
p=\frac{1}{2}$ as follow,

\begin{equation}
\frac{1}{2\left\vert V\right\vert N(0)}=\int_{0}^{\epsilon _{D}}\dfrac{%
d\zeta }{\sqrt{\zeta ^{2}+\Delta ^{2}}}\left[ \frac{1}{\left( 1+e^{\dfrac{-2%
\sqrt{\zeta ^{2}+\Delta ^{2}}}{k_{B}T}}\right) ^{\dfrac{1}{2}}}-\frac{1}{%
\left( 1+e^{\dfrac{2\sqrt{\zeta ^{2}+\Delta ^{2}}}{k_{B}T}}\right) ^{\dfrac{1%
}{2}}}\right] \text{.}
\end{equation}

\begin{center}
Determine $T_{C}$
\end{center}

At $\Delta =0$, we have $T=T_{C}$, so from equations (32,33) we can write,

\begin{equation}
\frac{1}{2\left\vert V\right\vert N(0)}=\int_{0}^{\dfrac{\epsilon _{D}}{%
K_{B}T_{C}}}\dfrac{dx}{x}\left[ \frac{1}{\left( 1+e^{-2x}\right) ^{\dfrac{1}{%
2}}}-\frac{1}{\left( 1+e^{2x}\right) ^{\dfrac{1}{2}}}\right] \text{.}
\end{equation}

Define $\phi (x)=\left[ \dfrac{1}{\left( 1+e^{-2x}\right) ^{\dfrac{1}{2}}}-%
\dfrac{1}{\left( 1+e^{2x}\right) ^{\dfrac{1}{2}}}\right] $, we can calculate
the last integral as,

\begin{equation}
\int_{0}^{\dfrac{\epsilon _{D}}{K_{B}T_{C}}}\dfrac{dx}{x}\phi (x)\approx %
\left[ \phi (x)\ln x\right] _{0}^{\dfrac{\epsilon _{D}}{K_{B}T_{C}}%
}-\int_{0}^{\infty }dx\dfrac{d\phi (x)}{x}\ln x=\ln \dfrac{\epsilon _{D}}{%
K_{B}T_{C}}+0.4228=\ln \dfrac{1.535\epsilon _{D}}{K_{B}T_{C}}\text{.}
\end{equation}

So,we can write the equation(34) as,

\begin{equation}
\frac{1}{2\left\vert V\right\vert N(0)}=\ln \dfrac{1.535\epsilon _{D}}{%
K_{B}T_{C}}\text{,}
\end{equation}

or on the equivalent form,

\begin{equation}
T_{C}=\dfrac{1.535\epsilon _{D}}{K_{B}}e^{\dfrac{-1}{2\left\vert
V\right\vert N(0)}}\text{.}
\end{equation}

Comparing equation (37), the critical temperature for the Anyon case ($P=%
\frac{1}{2}$) and equation (13), the critical temperature for BCS theory (at
the same value of Debye energy) we have,

\begin{equation}
\frac{T_{C_{{\huge p=}\frac{1}{2}}}}{T_{C_{\text{BCS}}}}=1.35e^{\dfrac{1}{%
2\left\vert V\right\vert N(0)}}\text{.}
\end{equation}

So, for some weak coupling limit, say $\dfrac{1}{\left\vert V\right\vert N(0)%
}=5$, we have

\begin{equation}
T_{C_{{\huge p=}\frac{1}{2}}}\simeq 16,446T_{C_{\text{BCS}}}\text{.}
\end{equation}

So, if the critical temperature in the standard BSC is equal to $1$ Kelvin
(with $\dfrac{1}{\left\vert V\right\vert N(0)}=5$), then using a unified
statistics with $P=\frac{1}{2}$, the critical temperature jumps to $16,446$
Kelvin.

Notice that although the weak coupling is the reason of why we cannot
explain some HTS using the standard BCS theory, in the Anyonic statistics
the critical temperature is increased whenever the coupling limit is
decreased!!

As an example, consider that we have more weak interaction (say $\dfrac{1}{%
\left\vert V\right\vert N(0)}=10$), then we have,

\begin{equation}
T_{C_{{\huge p=}\frac{1}{2}}}\simeq 201T_{C_{\text{BCS}}}!!
\end{equation}

So at the same condition, if the critical temperature in the standard BSC is
eual to $1$ Kelvin then for a weak coupling $\dfrac{1}{\left\vert
V\right\vert N(0)}=5$ we see that the critical temperature jumps to $16,446$
Kelvin and for more weak coupling $\dfrac{1}{\left\vert V\right\vert N(0)}%
=10 $ the critical temperature jumps to $201$ Kelvin!

\begin{center}
\textbf{Determine }$\Delta _{0}$
\end{center}

At $T=0$, we have $\Delta =\Delta _{0}$, so from the equation (29) we have,

\begin{equation}
\frac{1}{2\left\vert V\right\vert N(0)}=\int_{0}^{\epsilon _{D}}\dfrac{%
d\zeta }{\sqrt{\zeta ^{2}+\Delta _{0}^{2}}}=\sinh ^{-1}\left( \frac{\epsilon
_{D}}{\Delta _{0}}\right) \text{ ,}
\end{equation}

so we have,

\begin{equation}
\Delta _{0}=2\epsilon _{D\text{ }}e^{\dfrac{-1}{2\left\vert V\right\vert N(0)%
}}\text{..}
\end{equation}

From (24) and (28), we have

\begin{equation}
\Delta _{0}=1.3K_{B}T_{C}\text{.}
\end{equation}

Finally, we can plot the universal curve $y=y(x)$, where $x=\dfrac{T}{T_{C}}%
,y=\dfrac{\Delta }{\Delta _{0}}$ for $p=\frac{1}{2}$ as follow,

\begin{equation}
\frac{1}{2\left\vert V\right\vert N(0)}=\int_{0}^{\dfrac{\sinh \frac{1}{%
2\left\vert V\right\vert N(0)}}{y}}\dfrac{dz}{\sqrt{z^{2}+1}}\left[ \frac{1}{%
\left( 1+e^{\dfrac{-2.6y\sqrt{z^{2}+1}}{x}}\right) ^{\dfrac{1}{2}}}-\frac{1}{%
\left( 1+e^{\dfrac{2.6y\sqrt{z^{2}+1}}{x}}\right) ^{\dfrac{1}{2}}}\right] 
\text{.}
\end{equation}

\subsection{$p=\frac{1}{3}$ Anyon superconductor}

For $p=\frac{1}{3}$, we can determine $\phi (r)$ from equation (21) as,

\begin{equation}
\phi (r)=\frac{3}{(2r^{3}+2\sqrt{r^{6}+r^{3}}+1)^{\dfrac{1}{3}}+(2r^{3}+2%
\sqrt{r^{6}+r^{3}}+1)^{\dfrac{-1}{3}}-1}\text{.}
\end{equation}

From equations (22-23) and after some calculations, we can find the formula
of $T_{C}$ for the Anyon $p=\frac{1}{3}$ as,

\begin{equation}
T_{C}=\frac{32.627}{K_{B}}\epsilon _{D}\text{ }e^{\dfrac{-1}{3\left\vert
V\right\vert N(0)}}\text{.}
\end{equation}

Also, from the general equation (31) we have,

\begin{equation}
\Delta _{0}=2\epsilon _{D}\text{ }e^{\dfrac{-1}{3\left\vert V\right\vert N(0)%
}}\text{.}
\end{equation}

\begin{conclusion}
In this paper, we calculated the critical temperature of some metals using
the standard BCS theory but assuming that the particles inside this metals
obey Anyonic statistics instead Fermi-Dirac statistics. We find that the
lower \ of the interaction we have, the greater of the critical temperature
which may be a reasonable explanation of some high temperature
superconductors which the standard BCS theory cannot explain it.

\begin{acknowledgement}
I want to thank E. Ahmad for his advice and very useful discussion.
\end{acknowledgement}
\end{conclusion}

\section{References}

[1] Simon Reif-Acherman, "Heike Kamerlingh Onnes: Master of Experimental
Technique and Quantitative Research", \textit{Physics in Perspective, }vol. 
\textbf{6}, pp. 197--223, (2004).

[2] W. Meissner, R. Ochsenfeld, "Ein neuer Effekt bei Eintritt der
Supraleitfahigkeit", \textit{Naturwissenschaften}, vol. \textbf{21}, pp.
787--788, (1933).

[3] J. Bardeen, L. N. Cooper, J. R. Schrieffer, "Microscopic Theory of
Superconductivity", \textit{Physical Review}, vol. \textbf{106}, (1957).

[4] C. S. Gorter, H.Casimir, "On superconductivity", \textit{Zeitschrift f%
\"{u}r Technische Physik , }vol. \textbf{15}, (1934).

[5] H. London and F. London, "The Electromagnetic Equations of the
Supraconductor", \textit{Proceedings of the Royal Society A}, vol.\textbf{149%
}, (1935).

[6] V. L. Ginzburg, L. D. Landau, \textit{Soviet Physics JETP }20, vol. 
\textbf{1064}, (1950).

[7] J. G. Bednorz, K. A. Muller, "Possible high TC superconductivity in the
Ba-La-Cu-O system", \textit{Zeitschrift f\"{u}r Physik B}, vol. \textbf{64, }%
pp. 189--193, (1986).

[8] A. A. Abutaleb, "Unified statistical distribution of quantum particles
and Symmetry",\textit{\ Int. J. Theor. Phys}, vol. \textbf{53, }pp\textbf{. }%
3893-3900, (2014).

[9] http://www-personal.umich.edu/\symbol{126}sunkai/teaching/Fall%
\_2012/phys620.html.

\end{document}